# A Novel Second-Order Nonlinear Differentiator With Application to Active Disturbance Rejection Control

Ibraheem Kasim Ibraheem

Department of Electrical Engineering, College of Engineering, University of Baghdad
Baghdad, Iraq
ibraheemki@coeng.uobaghdad.edu.iq

Wameedh Riyadh Bdul-Adheem

Department of Electrical Engineering, College of Engineering University of Baghdad
Baghdad, Iraq
wameedh.r@coeng.uobaghdad.edu.iq

*Abstract*—A Second-order Nonlinear Differentiator (SOND) is presented in this paper. By combining both linear and nonlinear terms, this tracking differentiator shows better dynamical performances than other conventional differentiators do. The hyperbolic tangent tanh(.) function is introduced due to two reasons; firstly, the high slope of the continuous tanh(.) function near the origin significantly accelerates the convergence of the proposed tracking differentiator and reduces the chattering phenomenon. Secondly, the saturation feature of the function due to its nonlinearity increases the robustness against the noise components in the signal. The stability of the suggested tracking differentiator is proven based on the Lyapunov analysis. In addition, a frequency-based analysis is applied to investigate the dynamical performances. The performance of the proposed tracking differentiator has been tested in active disturbance rejection control (ADRC) paradigm, which is a recent robust control technique. The numerical simulations emphasize the expected improvements.

## I. INTRODUCTION

One of the practical problems in the control engineering is the differentiation of signals. The presence of noise within the signal to be differentiated will cause distortion in the derivative of that signal [1]. Therefore, it is a requirement to design a differentiator that provides a derivative of an input signal even if a high-frequency noise has been detected within the signal to be differentiated [2]. Aiming a differentiator as a distinct module is a familiar design purpose in the area of control engineering. The preliminary method is to allow a linear dynamic model to represent the differentiator in question. Hence, the designed tracking differentiator will not estimate the exact derivatives due to bandwidth limitations [2]. In the last twenty years, tracking differentiator (TD) has a considerable attention due to its use as an essential part in the field of navigation and control systems [3]. The classical "high gain" differentiators stated in [4] may possibly track the exact derivatives when the gains approach infinity which is practically unrealizable. In [2], a new tracking differentiator has been designed based on sliding mode technique, the Lipschitz constant of the signal has to be upper bounded to achieve exact differentiation in this type of differentiator. Due to the presence of a discontinuous function, the estimated derivative is not smooth and the chattering phenomenon arises in this class of differentiators. An exact differentiator was established in [5] by integrating the behavioral principles of the sliding mode and the "high gain" differentiators with the use of a switching function. A continuous hybridized nonlinear differentiator is proposed in [6,7] in which the chattering is reduced. A finite time exact convergence is suggested in [8, 9]. In practice, to accomplish extreme performance, several applications based on tracking differentiators have been recommended, for example, in the control of underwater vehicles [10], in single-phase active power filters the TD is used for detecting the harmonic currents [11]. Additionally, the success has been established by many engineering systems [12]–[19].

The contribution of this paper is the construction of a second order nonlinear tracking differentiator, which includes a hyperbolic parameterized tangent function that improves the dynamical performance through significant slope around the origin. Additionally, the proposed differentiator is robustly capable to handle noisy versions of the input signal, which in turn produces a noise-free 1st derivative of the signal.

The paper is organized as follows: section II includes the problem statement. Next, in section III an improved nonlinear tracking differentiator is presented and a mathematical model is completely described. In section IV, a theoretical foundation is established, which is followed by a numerical simulation and results discussion in section V. Conclusion and remarks are given in section VI.

## II. PROBLEM STATEMENT

The differentiator, in principle, is an estimator, which is model-independent. Given a signal $v(t)$, the real-time differentiation difficult involved in obtaining an estimate of its derivative $\dot{v}(t)$. The differentiator is expected to be constructed the following form:

$$\dot{x}_1(t) = x_2(t) \ , \quad \dot{x}_2(t) = f(x_1(t) - r(t), x_2(t))$$

Assuming there a solution to the above dynamical system, a differentiator can be designed with the following being fulfilled, $x_1(t)$ tends to $r(t)$ and $x_2(t)$ tends to $\dot{r}(t)$.

## III. THE PROPOSED SECOND-ORDER NONLINEAR DIFFERENTIATOR (SOND)

The nonlinear second order differentiator proposed in this work utilizing the hyperbolic tangent function,

$$\begin{cases} \dot{x}_1(t) = x_2(t) \\ \dot{x}_2(t) = -\rho^2 \tanh\left(\dfrac{bx_1(t) - (1-a)r}{c}\right) - \rho x_2(t) \end{cases} \quad (1)$$

Where $x_1(t)$ is an estimation of the actual input $r$, and $x_2(t)$ is an estimation of the derivative of the actual input. the coefficients $a, b, c,$ and $\rho$ are design factors, where $0 < a < 1, b > 0, c > 0,$ and $\rho > 0$.

## IV. THEORETICAL ASPECTS OF THE PROPOSED DIFFERENTIATOR

**Lemma 1 :( The Convergence of the SOND):** The Dynamical System **represented by (1)** is globally asymptotically stable.

**Proof:** Assume that $V_l(x) = \rho \frac{\gamma}{b} \ln \cosh\left(\frac{bx_1}{c}\right) + \frac{1}{2} x_2^2$ is candidate Lyapunov function to dynamical the system (1). In this case, $V_l(x) > 0$ iff $x \neq 0$, while for $x = 0$, $V_l(x) = 0$. Now, $\dot{V}_l(x) = -Rx_2^2(t)$ and $\dot{V}_l(x) \leq 0$ for all $x_2(t)$ and $\dot{V}_l(0) = 0$ at the origin by "LaSalle theorem" [20], because for $\|x(t)\| \to \infty$, $V_l(x) \to \infty$, then, the system is asymptotically stable in the global sense (GAS). □

Another method to prove the asymptotic stability is given in Lemma 3 given next.

**Lemma 2: (Phase of Arrival):** for the dynamical system described by (1), if $\frac{bx_1(t)-(1-a)r}{c} \gg 1$; then $\forall t > 0$, $\frac{bx_1(t)-(1-a)r}{c}$ is a decreasing function of time where it approaches the tracking phase at which $\left|\frac{bx_1(t)-(1-a)r}{c}\right| < \varepsilon$.

**Proof:** given $\frac{bx_1(t)-(1-a)r}{c} \gg 1$, Then $\tanh\left(\frac{bx_1(t)-(1-a)r}{c}\right) \to 1$, so that:

$$\begin{cases} \dot{x}_1(t) = x_2(t) \\ \dot{x}_2(t) = -\rho^2 - \rho x_2(t) \end{cases} \quad (2)$$

Assuming the initial condition $x(0) = [x_1(0) \quad x_2(0)]^T$, the above dynamical systems (2) is set of $1^{st}$ order ordinary differential equations with a solution in the time-domain given as,

$$x_1(t) = 1 + x_1(0) + \frac{x_2(0)}{\rho} + -\rho t - \left(1 + \frac{x_2(0)}{\rho}\right)e^{-\rho t}$$

$$x_2(t) = -\rho + (\rho + x_2(0))e^{-\rho t}$$

Apparently, $x_1(t)$ is a decreasing function in terms of $t$ for $t \in [0, T]$, where $T$ is the time, at which the tracking differentiator enters the Phase of tracking where $\left|\frac{bx_1(t)-(1-a)r}{c}\right| < \varepsilon$. □

**Corollar1:** Consider the tracking differentiator represented according to (1). If $\frac{bx_1(t)-(1-a)r(t)}{c} \ll -1$, then $\forall t > 0$, $\frac{bx_1(t)-(1-a)r(t)}{c}$ is an increasing w.r.t $t$ and at certain point of time, the tracking differentiator enters the Phase of tracking, i.e. where $\left|\frac{bx_1(t)-(1-a)r}{c}\right| < \varepsilon$.

**Proof:** Following the proof in Lemma (2), $x_1(t)$ is a decreasing function in terms of $t$ for $t \in [0, T]$, where $T$ is the time at which the tracking differentiator enters the Phase of tracking where $\left|\frac{bx_1(t)-(1-a)r}{c}\right| < \varepsilon$. □

**Lemma 3(Phase of tracking):** Consider the dynamical system represented by (1), if $\left|\frac{bx_1(t)-(1-a)r}{c}\right| < \varepsilon$, then both tracking error $e_t(t) = r(t) - \frac{b}{1-a} x_1(t)$, and the differentiation error $e_d(t) = \dot{r}(t) - \frac{b}{1-a} x_2(t)$ approach zero for bounded input signal.

**Proof:** due to the fact that $\frac{bx_1(t)-(1-a)r}{c} < \varepsilon$, then $\tanh\left(\frac{bx_1(t)-(1-a)r}{c}\right) \to \left(\frac{bx_1(t)-(1-a)r}{c}\right)$. So that,

$$\begin{cases} \dot{x}_1(t) = x_2(t) \\ \dot{x}_2(t) = -\rho^2 \left(\dfrac{bx_1(t) - (1-a)r}{c}\right) - \rho x_2(t) \end{cases} \quad (3)$$

taking the Laplace transform to (3), we get

$$\begin{bmatrix} X_1(s) \\ X_2(s) \end{bmatrix} = \begin{bmatrix} \dfrac{\rho^2(1-a)}{c} & \dfrac{\rho^2(1-a)s}{c} \\ s^2+\rho s+\dfrac{\rho^2 b}{c} & s^2+\rho s+\dfrac{\rho^2 b}{c} \end{bmatrix}^T R(s) \quad (4)$$

The characteristic equation of transfer function (4) is $s^2 + \rho s + \frac{\rho^2 b}{c}$ with roots, $s_{1,2} = -\frac{\rho}{2} \mp \sqrt{\frac{\rho^2}{4} - \frac{\rho^2 b}{c}}$, hence the proposed tracking differentiator is GAS. During the Phase of tracking, the tracking error can be expressed as is $e_t(t) = r(t) - \frac{b}{1-a} x_1(t)$ and in Laplace-domain is given as $E_t(s) = R(s) - \frac{b}{1-a} X_1(s)$. The following relation, expresses the tracking error in terms of the input of the racking differentiator,

$$L_t(s) = \frac{E_t(s)}{R(s)} = \frac{s(s+\rho)}{s^2 + \rho s + \dfrac{\rho^2 b}{c}}$$

So that,

$$l_t(\infty) = \lim_{s \to 0} sL_t(s) = 0 \quad (5)$$

While the differentiation error through tracking phase is given as, $e_d(t) = \dot{r}(t) - \frac{b}{1-a} X_2(t)$, and in Laplace-domain $E_d(s) = sV(s) - \frac{b}{1-a} X_2(s)$. The relationship between the differentiation error and the input derivative is given as $L_d(s) = \frac{E_d(s)}{sR(s)} = \frac{s(s+\rho)}{s^2+\rho s+\frac{\rho^2 b}{c}}$. So that,

$$l_d(\infty) = \lim_{s \to 0} sL_d(s) = 0 \quad (6)$$

Hence, (5) and (6) finishes the proof. □

**Theorem 1:** Suppose that we have the dynamical system given by (1), then whatever the value of $\left|\frac{bx_1-(1-a)r}{c}\right|$, $\lim_{t \to \infty} \left|\frac{bx_1(t)-(1-a)r(t)}{c}\right| = 0$ and $\lim_{t \to \infty} \left|\frac{bx_2(t)-(1-a)\dot{r}(t)}{c}\right| = 0$.

**Proof:** Using Lemma (2) and (3). □

**Lemma 4:** Suppose the dynamical system described by (1) and satisfying the Phase of tracking conditions, i.e. (3). If $b \gg 1, 0 < c < 1, \rho \gg 1,$ with $0 < a < 1$, then dynamical system described by (1) has a high value of $\omega_n$, a small value of $\xi$, and a peaking phenomenon.

**Proof:** it is given that $\frac{x_2(s)}{sR(s)} = \frac{\frac{\rho^2(1-a)}{c}}{s^2+\rho s+\frac{\rho^2 b}{\gamma}} = \left(\frac{1-a}{b}\right)\frac{\omega_n^2}{s^2+2\xi\omega_n s+\omega_n^2}$, where, $\omega_n = \rho\sqrt{\frac{b}{c}}$ in (rad/sec) and $\xi = \frac{1}{2}\sqrt{\frac{c}{b}}$. Given $b, c, and\ \rho$, hence $\xi \ll 1$ infers that the tracking differentiator written expressed in (1) has an "under-damped-effect" leading to "peaking phenomenon". □

**Lemma 5:** Assume that we have the dynamical system denoted by (1) which fulfills system represented by (3) with its coefficients $b$, $c$, and $\rho$ introduced previously in Lemma 4. The tracking differentiator is a band-limiting differentiator for $\omega < \omega_n$.

**Proof:** Given $\frac{x_2(j\omega)}{R(j\omega)} = \left(\frac{1-a}{b}\right)\frac{\omega_n^2\ j\omega}{(j\omega)^2+2\xi\omega_n j\omega+\omega_n^2} = \left(\frac{1-a}{b}\right)\frac{j\omega}{\left(\frac{j\omega}{\omega_n}\right)^2+2\xi\frac{j\omega}{\omega_n}+1}$, expressing this transfer function in logarithmic scale as $20*\log\left|\frac{x_2(j\omega)}{V(j\omega)}\right| = 20.\log\left(\frac{1-a}{b}\right) + 20.\log\omega - 20.\log\sqrt{(1-\left(\frac{\omega}{\omega_n}\right)^2)^2 + \left(2\xi\frac{\omega}{\omega_n}\right)^2}$. Now for $\omega \ll \omega_n$, this entails $20.\log\left|\frac{x_2(j\omega)}{R(j\omega)}\right| = 20.\log\left(\frac{1-a}{b}\right) + 20.\log\omega$, so, $20.\log\left(\frac{1-a}{b}\right)$ is the "correction-gain" and $20\log\omega$ is the "differentiator effect". On the other hand for $\omega \gg \omega_n$. Then $20.\log\left|\frac{X_2(j\omega)}{R(j\omega)}\right| = 20.\log\left(\frac{1-a}{b}\right) + 20.\log\omega - 40.\log\frac{\omega}{\omega_n}$, the term $40\log\frac{\omega}{\omega_n}$ represents the "attenuation-effect". Hence, the proposed nonlinear differentiator has an attenuation effect for $\omega \gg \omega_n$. □

## V. APPLICATION OF THE SOND IN ACTIVE DISTURBANCE

The classical ADRC proposed by *J. Han* [15] is built by combining the tracking differentiator (TD), the nonlinear state error combination (NLSEF), and the linear extended state observer (LESO) [22]. In Fig. 1, the improved version of the ADRC (IADRC) is illustrated.

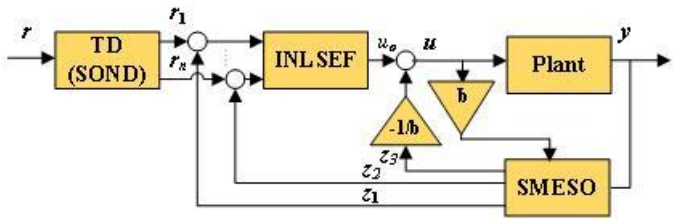

Fig.1. The IADRC topology.

In the INLSEF controller, the algorithm uses the *sign*(.) together with the exponential function which are integrated as follows, $u_{INLSEF} = \Psi(e) = k(e)^T f(e) + u_{int}$, Where $e \in \mathbb{R}^n$ is the vector of the state error, defined as $e = [e^{(0)}\ ....e^{(i)}\ ....\ e^{(n-1)}]^T$ [23]. In this regard, $e^{(i)}$ is the *i-th* derivative of the state error defined as, $e^{(i)} = x^{(i)} - z^{(i)}$. The function $k(e)$ is the nonlinear gain, defined as:

$$k(e) = \begin{bmatrix} k(e)_1 \\ \vdots \\ k(e)_i \\ \vdots \\ k(e)_n \end{bmatrix} = \begin{bmatrix} \left(k_{11} + \frac{k_{12}}{1+exp(\mu_1(e^{(0)})^2)}\right) \\ \vdots \\ \left(k_{i1} + \frac{k_{i2}}{1+exp(\mu_n(e^{(i-1)})^2)}\right) \\ \vdots \\ \left(k_{n1} + \frac{k_{n2}}{1+exp(\mu_n(e^{(n-1)})^2)}\right) \end{bmatrix} \quad (7)$$

The coefficients $k_{i1}, k_{i2}, \mu_i \in \mathbb{R}^+$ are the controller design parameters. The function $f(e)$ is the error function, defined as:

$$f(e) = \left[|e^{(0)}|^{\alpha_1}sign(e) \quad ...|e^{(i)}|^{\alpha_i}sign(e^{(i)}) \quad ......\right.$$

$$\left....\ |e^{(n-1)}|^{\alpha_n}sign(e^{(n)})\right]^T \quad (8)$$

$$u_{int} = (|\int e\ dt|^\alpha sign(\int e\ dt))\frac{k}{1+exp(\mu(\int e\ dt)^2)} \quad (9)$$

Finally, the nominal control signal $u_0 = \delta \tanh(\frac{u_{integrator}}{\delta})$. The SMESO (for $n$ = 2) has the following state-space representation [24]:

$$\begin{cases} \dot{z}_1 = z_2 + \beta_1(K_\alpha|y-z_1|^\alpha sign(y-z_1) + \\ \qquad K_\beta|y-z_1|^\beta(y-z_1)) \\ \dot{z}_2 = z_3 + bu + \beta_2(K_\alpha|y-z_1|^\alpha sign(y-z_1) + \\ \qquad K_\beta|y-z_1|^\beta(y-z_1)) \\ \dot{z}_3 = \beta_3(K_\alpha|y-z_1|^\alpha sign(y-z_1) + \\ \qquad K_\beta|y-z_1|^\beta(y-z_1)) \end{cases} \quad (10)$$

where $\mathbf{z} = [z_1, z_2, z_3]^T \in \mathbb{R}^3$, is a vector that includes the predictable states of the plant and the total-disturbance. The coefficients $\beta_i(i = 1,2,3), K_\alpha, \alpha, K_\beta$ and $\beta \in \mathbb{R}^+$ are SMESO design parameters.

## VI. NUMERICAL SIMULATION

In this section, some numerical simulations are carried out for the proposed second-order tracking differentiators given in (1). The numerical code is programmed in MATLAB®. The simulation was conducted by the $4^{th}$ order Runge-Kutta method with step size equals to 0.002. It was taken that $t_0 = 0$ and $t_f = 2$sec, initial values of the internal variable $\boldsymbol{x}(0)$ is zero. This numerical simulation includes comparing the proposed nonlinear differentiator (1), with five differentiators which are: high gain differentiator (HGTD)[4], robust exact differentiator (RED)[2], hybrid continuous nonlinear differentiator (HCND)[6], rapid convergent nonlinear differentiator (RCND) [7], and robust exact uniformly convergent arbitrary order differentiator (REUCAOD) [8][9][2]. A differentiator was tested using the signal $\sin(2\pi t) + n(t)$ as the input signal $r(t)$, a free analog signal which is supposed to be measured in continuous manner. In the following simulations, The noise component $n(t)$ is considered to have the following two cases:

1. Derivative estimation with low-frequency noise component with magnitude 0.001
$$n(t) = 0.001\sin(2\pi * 10t)$$
2. Derivative estimation with high-frequency noise component with magnitude 0.1
$$n(t) = 0.1\sin(2\pi * 16000t).$$

The four indices that are used to illustrate the performance of the proposed tracking differentiator are defined in [21], these are MSE, IAE, ITAE, ITSE., where the measured error is defined as, $e(t) = 2\pi \cos(2\pi t) - z_2$. The results of the first case with $v(t) = \sin(2\pi t) + 0.001\sin(2\pi * 10t))$ are shown in Figs. 2-7. The selected parameters and the numerical quantities are presented in table I.

TABLE I. THE TRACKING DIFFERENTIATORS PARAMETERS.

| Differentiator | Parameters |
|---|---|
| HGTD | $a_1=1.5990$, $a_2=280.8875$, $\tau=0.0111$ |
| RED | $C_2=39.4784$, $\lambda_1=9.4248$, $\lambda_2=43.4263$ |
| HCND | $k_1=0.5$, $k_2=150$, $k_3=765$, $k_4=150$, $\alpha=0.55$ |
| RCND | $\varepsilon=0.10857$, $\alpha=0.85077$, $a_{10}=122.1329$, $a_{11}=3.44665$, $a_{20}=0.073733$, $a_{21}=0.653865$ |
| REUCAOD | $c_2 = 39.4784$, $k_1 = 3.1416$, $k_2 = 3754.4$, $\kappa_1=120.05098$, $\kappa_2=119.31439$, $\alpha=0.00604$, $T_u=0.4$ |
| Proposed SOND | $a = 0.958128$, $b = 128.13044$, $c = 0.03758384$, $\rho = 19.159814$ |

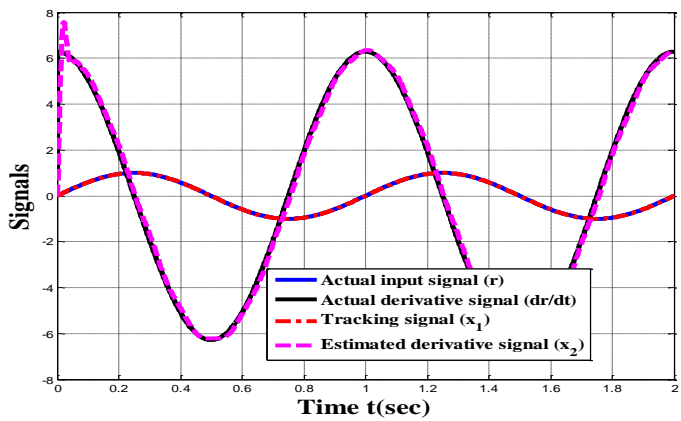

Fig. 2. HGTD tracking $r$ ($t$) wave.

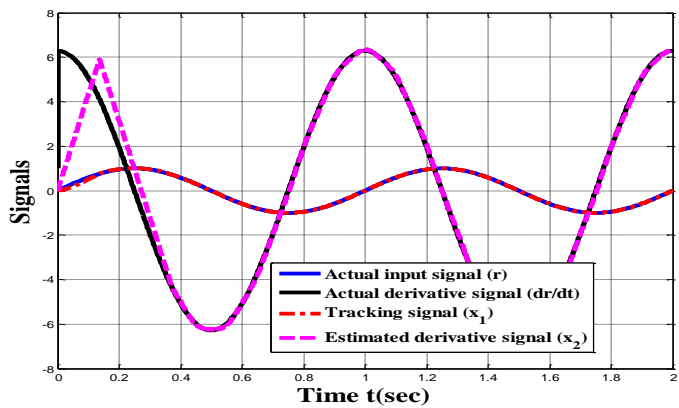

Fig. 3. RED tracking $r$ ($t$) wave.

The parameters of the tracking differentiators are listed in table I. The measurement indices that are carried out from this test are shown in table II.

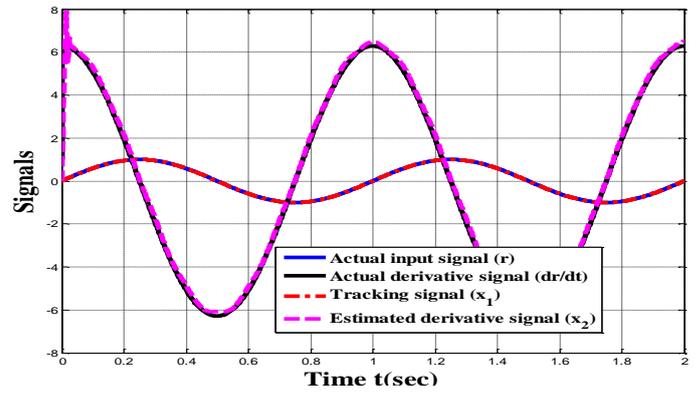

Fig. 4. HCND tracking $r$ ($t$) wave.

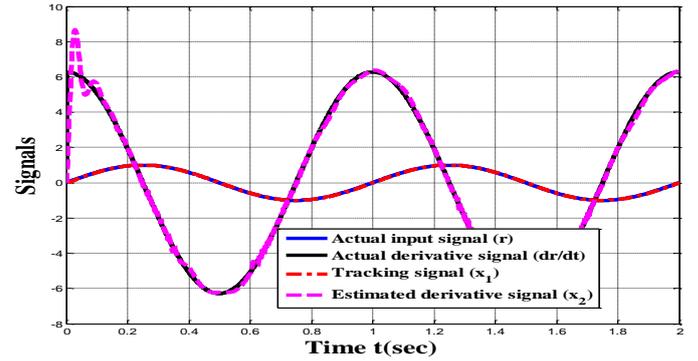

Fig. 5. RCND tracking $r$ ($t$) wave.

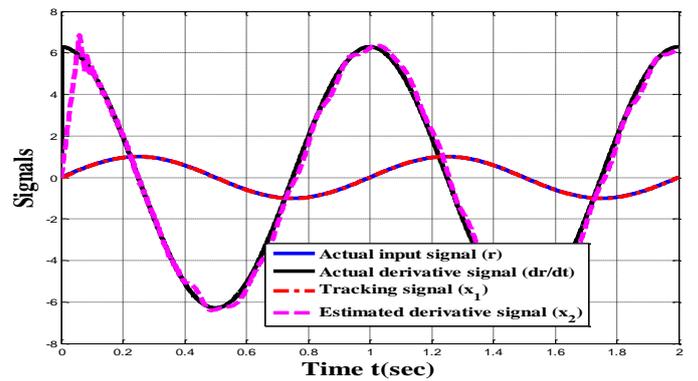

Fig. 6. REUCAOD tracking $r$ ($t$) wave.

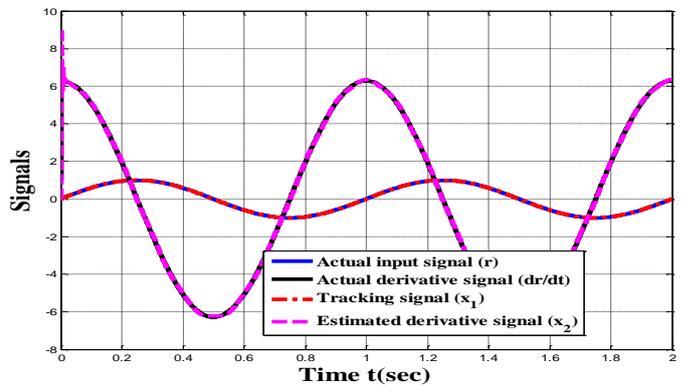

Fig. 7. SOND tracking $r$ ($t$) wave of case 1.

For the second test with the high-frequency noise component, where $r(t) = \sin(2\pi t) + 0.1 \sin(2\pi * 16000 t))$. An improvement in the quality of the SOND has been obtained. Tables II and III shows the improvement in the performance of the SOND. Fig. 8 represents the resulting tracking to the input signal $r(t)$,

TABLE II. PERFORMANCE MEASURES OF CASE 1

| Differentiator | MSE | IAE | ITAE | ITSE |
|---|---|---|---|---|
| HGTD | 0.134426 | 0.357909 | 0.294527 | 0.056701 |
| RED | 0.944760 | 0.705011 | 0.173515 | 0.111786 |
| HCND | 0.083079 | 0.290542 | 0.239004 | 0.034810 |
| RCND | 0.182127 | 0.370517 | 0.256905 | 0.047512 |
| REUCAOD | 0.402831 | 0.591486 | 0.479002 | 0.165063 |
| SOND | 0.011647 | 0.091862 | 0.081136 | 0.004204 |

TABLE III. PERFORMANCE MEASURES OF CASE 2

| Differentiator | MSE | IAE | ITAE | ITSE |
|---|---|---|---|---|
| HGTD | 0.132042 | 0.350477 | 0.287009 | 0.051909 |
| RED | 0.930758 | 0.672875 | 0.149013 | 0.102578 |
| HCND | 0.081205 | 0.291592 | 0.239325 | 0.030860 |
| RCND | 0.175575 | 0.335618 | 0.225080 | 0.039785 |
| REUCAOD | 0.392482 | 0.582985 | 0.468142 | 0.138247 |
| SOND | 0.009227 | 0.029341 | 0.017900 | 0.000239 |

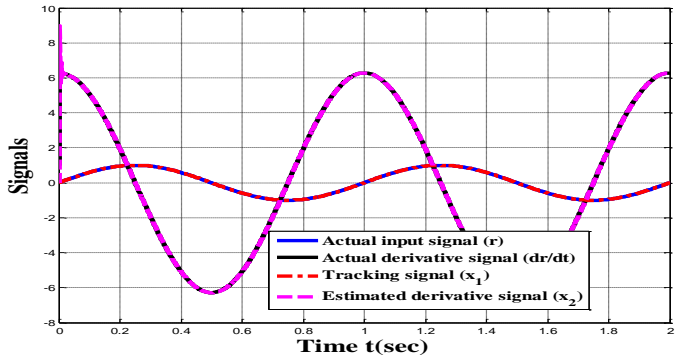

Fig. 8. SOND tracking $r(t)$ wave of case 2.

From the simulations, it is shown that the chattering phenomenon is alleviated adequately, and swift and high accuracy tracking is assured. Unfortunately, the peaking phenomenon appears as an essential part of the follow-up response, but it can be reduced by considering system (1) with a different set of parameters. This useful technique permits us to treat the trade-off between the differentiation error, and the peaking rate of the derivative estimates. In addition, it is cleared that as the frequency of the $n(t)$ noise component added to the original signal is increased significantly for 5Hz to 8KHz, which is associated with rising in its amplitude from 0.001 to 0.1, the SOND shows an insignificant change in the MSE performance index. This minor change is due to the overshoot phenomenon that is related to the parameters of the proposed differentiator. But, the significant improvement can be shown clearly in the other indices with reduction by 68.1% for IAE, 80% for ITAE, and 94% for ITSE indices. All of The parameters of the SOND (*a, b, c,* and $\rho$) have direct effect on its output response, e.g., increasing $\rho$ leads to an increase of the SOND's bandwidth($\omega_n$). Numerically, for the given values of the SOND parameters given in table I, $\omega_n = \rho(\frac{b}{c})^{1/2} = 1120$ rad/sec, which is the frequency after it all of the noise frequencies in the input signal will be attenuated by the SOND.

As an application of the SOND, the following numerical simulation includes the control of PMDC motor by using IADRC. The simplified nonlinear state space representation of the permanent magnet DC motor (PMDC) is given by[23]:

$$\begin{cases} \dot{x}_1 = x_2 \\ \dot{x}_2 = -\frac{R_a B_{eq} + K_t K_b}{L_a J_{eq}} x_1 - \frac{(L_a B_{eq} + R_a J_{eq})}{L_a J_{eq}} x_2 + \frac{1}{n}\frac{K_t}{L_a J_{eq}}(v_a + d) \\ y = x_1 \end{cases} \quad (10)$$

where $v_a$ is the input voltage applied to motor, $k_b$ is equal to the voltage constant, $k_t$ is the torque constant, $R_a$ is the electric resistance constant, $L_a$ is the electric self-inductance, $J_{eq}$ is the total-equivalent-inertia, $B_{eq}$ is the total-equivalent-damping of the combined motor rotor, gearbox, and load, $n$ is the gearbox ratio, and $T_L$ is the load torque applied at the shaft side. $x_1$, and $x_2$ are the angular speed and angular acceleration of the motor shaft. The equivalent disturbance at the input $d = \frac{L_a}{K_t}\dot{T}_L + \frac{R_a}{K_t}T_L$. The load torque $T_L = T_{ext} + F_c sgn(x_1)$ where $F_c$ is the coulomb friction force [25]. The values of the parameters for PMDC motor are [15] $R_a = 0.1557$, $L_a = 0.82$, $K_b = 1.185$, $K_t = 1.1882$, $n = 3.0$, $J_{eq} = 0.2752$, and $B_{eq} = 0.3922$. The parameters of the proposed INLSEF are $k_{11} = 144.2110$, $k_{12} = 4.7661$, $k_{21}=41.3437$, $k_{22}=2.3836$, $k_3=176.3737$, $\delta=8.8945$, $\mu_1=22.6214$, $\mu_2=29.4288$, $\mu_3=20.6845$, $\alpha_1 = 0.5940$, $\alpha_2 = 1.1272$, and $\alpha_3 = 5.6162$. The SOND proposed in this work has the following set of parameters: $a = 0.1055$, $b = 4.5528$, $c = 12.7228$, $\rho = 13.2749$. The parameters $K_\alpha = 0.7511$, $\alpha = 0.7490$, $K_\beta = 1.8629$, $\beta = 0.0331$, $\beta_1 = 19.403$, $\beta_2 =1084.9393$, $\beta_3 = 1880.1690$ represent the coefficients of the SMESO used in this work. The PMDC controlled by IADRC is tested by applying a reference angular- velocity equals to 1 rad per second at the time equals to zero ($t = 0$) and for $t = 10$ sec. To investigate the performance of the proposed IADRC, an external torque acting as a constant disturbance equal to 2 N.m is applied to the shaft during the simulation at $t = 5$ sec. The outputs of the SOND and SMESO with the tracking errors are shown in Fig. 9.

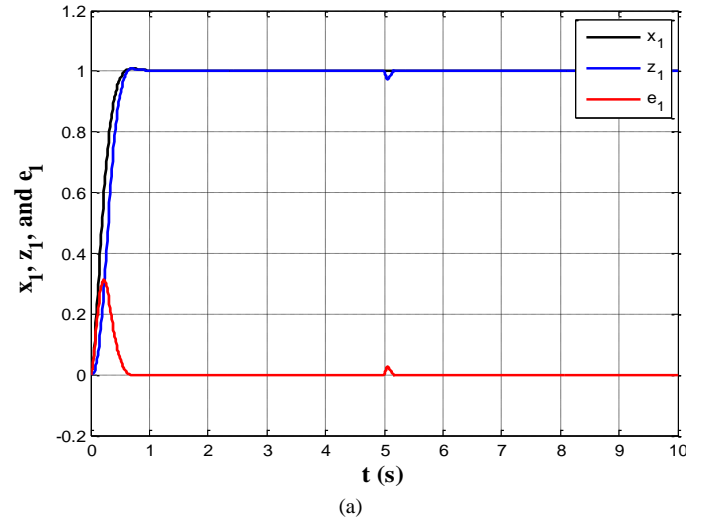

(a)

Figure 9, Continued…

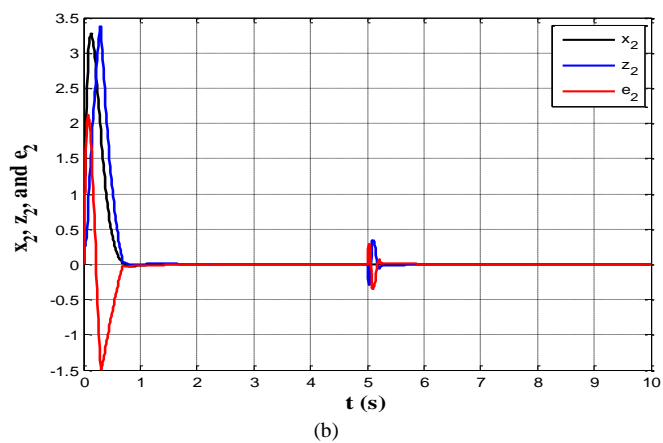

Fig. 9. The curves of the signals produced by the IADRC

## VII. Conclusions

In this paper, an improved nonlinear differentiator has been proposed based on hyperbolic tangent function. Based on the Lyapunov analysis, it has been proven that the SOND is a globally asymptotic stable. Simulations demonstrate that even with peaking phenomenon at the start of the tracking phase due to the underdamped effect, the SOND performs better than the other tracking differentiators in the comparisons in terms of the arrival time and MSE, ITAE, IAE, ITSE measures. Due to its continuous structure and as it encompasses of linear and nonlinear terms, the proposed SOND exhibits the band-limiting feature and suppresses the chattering phenomenon and noise components in the signal excellently as shown through the simulations. The performance of the SOND has been tested in ADRC paradigm and showed a very clear tracking performance. It enhances the precision of the differential estimation compared with the conventional method. A more precise approximate differentiation is achieved, which fulfills the higher requirements of the engineering applications specifically in the field of motion control.